\documentclass{book}    
\usepackage{piers,url}  
\newenvironment{sbmatrix}[1]{\left[ \begin{array}{#1}}%
	{\end{array} \right]}

\begin{document}

\title{A Nonlinear Boundary Condition for Continuum Models of Biomolecular Electrostatics}
\maketitle

\author      {J. P. Bardhan}
\affiliation {Northeastern University}
\city        {Boston, MA}
\postalcode  {02115}
\country     {USA}
\email       {j.bardhan@neu.edu}  
\nomakeauthor

\author      {D. A. Tejani}
\affiliation {Northeastern University}
\city        {Boston, MA}
\postalcode  {02115}
\country     {USA}

\author      {N. S. Wieckowski}
\affiliation {Northeastern University}
\city        {Boston, MA}
\postalcode  {02115}
\country     {USA}

\author      {A. Ramaswamy}
\affiliation {Northeastern University}
\city        {Boston, MA}
\postalcode  {02115}
\country     {USA}

\author      {M. G. Knepley}
\affiliation {Computation Institute, University of Chicago}
\city        {Chicago, IL}
\postalcode  {60637}
\country     {USA}
\email       {knepley@ci.uchicago.edu}  
\nomakeauthor

\begin{authors}

{\bf J. P. Bardhan}$^{1}$ {\bf (corresponding and presenting author)}, {\bf D. A. Tejani}$^{1}$, {\bf N. S. Wieckowski}$^{1}$, {\bf A. Ramaswamy}$^{1}$, {\bf and M. G. Knepley}$^{2}$\\
\medskip
$^{1}$Northeastern University, Boston, MA, USA (\texttt{j.bardhan@neu.edu})\\
$^{2}$University of Chicago, Chicago, IL, USA (\texttt{knepley@ci.uchicago.edu})

\end{authors}

\begin{paper}

\begin{piersabstract}
Understanding the behavior of biomolecules such as proteins requires
understanding the critical influence of the surrounding fluid
(solvent) environment---water with mobile salt ions such as sodium.
Unfortunately, for many studies, fully atomistic simulations of
biomolecules, surrounded by thousands of water molecules and ions are
too computationally slow.  Continuum-solvent models based on
macroscopic dielectric theory (e.g. the Poisson equation) are popular
alternatives, but their simplicity fails to capture well-known
phenomena of functional significance. For example, standard theories
predict that electrostatic response is symmetric with respect to the
sign of an atomic charge, even though response is in fact strongly
asymmetric if the charge is near the biomolecule surface. In this
work, we present an asymmetric continuum theory that captures the
essential physical mechanism---the finite size of solvent
atoms---using a nonlinear boundary condition (NLBC) at the dielectric
interface between the biomolecule and solvent.  Numerical calculations
using boundary-integral methods demonstrate that the new NLBC model
reproduces a wide range of results computed by more realistic, and
expensive, all-atom molecular-dynamics (MD) simulations in explicit
water.  We discuss model extensions such as modeling
dilute-electrolyte solvents with Debye-Huckel theory (the linearized
Poisson--Boltzmann equation) and opportunities for the
electromagnetics community to contribute to research in this important
area of molecular nanoscience and engineering.
\end{piersabstract}

\psection{INTRODUCTION}\label{sec:introduction}

Protein structure and function are determined in part by electrostatic
interactions between the protein's atomic charges and the surrounding
biological fluid (solvent), a complex mixture of polar water molecules
and dissolved charges such as sodium and potassium
ions~\cite{Sharp90}.  For over a century, biological scientists have
modeled these interactions using macroscopic continuum models based on
the Poisson--Boltzmann equation or Debye--H\"{u}ckel
theory~\cite{Sharp90,Bardhan12_review}.  These popular mean-field
theories assume that solvent molecules are infinitely small compared
to the biomolecule solute~\cite{Beglov96,Hildebrandt04}, a drastic
simplification critical to enable pioneering theoretical studies using
spherical and ellipsoidal models of protein shapes~\cite{Kirkwood34}.
However, its justifications are increasingly questionable for
atomistically detailed protein structures and predictions of binding
affinities (binding free energies), which are of enormous value to
understanding the molecular basis of disease and for designing
improved therapeutic drugs.

In contrast, exponential growth in computing capabilities has enabled
large-scale molecular-dynamics (MD) simulations that model the
surrounding solvent in fully atomistic, explicit
detail~\cite{Phillips05}.  These calculations can provide far more
realistic insights, but unfortunately require hundreds or thousands of
times the computational resources that continuum models need.  A
rigorous statistical-mechanical argument establishes the connection
between atomistic and continuum models~\cite{Beglov96,Roux99},
creating an opportunity to develop more accurate continuum models by
comparison to MD simulations, where new continuum theories can be
tested against a wide array of challenging ``computational
experiments'' which are unrealizable in real-world
laboratories~\cite{Ashbaugh00,Rajamani04,Mobley08,Bardhan12_asymmetry}.

We have been advancing multiple approaches to improve continuum
models~\cite{Bardhan11_pka,Bardhan12_asymmetry,Bardhan13_nonlocal_review,Bardhan14_asym}.
One approach replaces the traditional model of the solvent as a
macroscopic dielectric continuum (the familiar relation $D(r) =
\epsilon(r) E(r)$), and instead models it as a nonlocal dielectric
material~\cite{Dogonadze74,Hildebrandt04,Fedorov07}, which limits
short-range dielectric response using a convolution $D(r) =
\int_{\mathrm{solvent}} \epsilon(|r-r'|) E(r') dV'$.  On the other
hand, nonlocal models, and even many sophisticated nonlinear models,
are still symmetric with respect to the sign of the protein charges.
In reality, however, negative and positive charges near the solute
surface produce substantially different responses.  Asymmetry results
from both water structure at the interface, and the fact that water
hydrogens are much smaller than water
oxygens~\cite{Latimer39,Ashbaugh00,Rajamani04,Grossfield05,Corbeil10,Mukhopadhyay14,Bardhan12_asymmetry}.
We have demonstrated, using atomistic MD calculations, that
charge-sign asymmetries can be accurately reproduced if the electric
potential is modeled as a piecewise-linear plus constant
(i.e. piecewise-affine) function of the
charge~\cite{Bardhan12_asymmetry}.  The constant term in this model
represents the interface potential induced by water structure around a
completely uncharged version of the
solute~\cite{Cerutti07,Kathmann11}, and the discontinuity in response
coefficient occurs where the solute charge changes
sign~\cite{Bardhan12_asymmetry}.

The few theories that directly address hydration
asymmetry~\cite{Purisima09,Corbeil10,Mukhopadhyay14} are not actually
Poisson models, but generalizations of the Born-ion problem (the
spherically symmetric case of a sphere with a central charge).
Recently we proposed the first successful asymmetric Poisson theory
that can be solved for complex molecular
geometries~\cite{Bardhan14_asym}, by translating the existing models'
physical insights into a boundary-integral equation (BIE) formulation
of the Poisson problem~\cite{Altman09,Bardhan10}.  This led to a
modified BIE formulation in which we replaced the usual Maxwell
boundary condition for the continuity of the normal flux with a
nonlinear boundary condition (NLBC)~\cite{Bardhan14_asym}.
Calculations showed that the new model successfully reproduces highly
accurate MD calculations on a wide range of challenging test
cases~\cite{Mobley08}.

In the present paper, we improve the new NLBC model by ensuring that
the potential outside the solute protein satisfies Gauss's law
sufficiently far from the solute.  To demonstrate the new model, we
have implemented a boundary-element method solver in MATLAB (all of
the source code and data required to reproduce the figures in this
paper are freely and publicly available
online~\cite{PIERS15_repository}).  Calculations demonstrate that the
improved model is more accurate for monovalent atomic ions, and
illustrate that charge-sign hydration asymmetry effects are
substantial for surface charges, with the magnitude of asymmetry
decreasing rapidly for charges further from the solute--solvent
interface.  In particular, we predict that for surface charges in a
sphere, the energetic difference between positive and negative charges
does not depend strongly on the size of the sphere; this result
implies that asymmetry is an essential piece of physics to include in
models of molecular electrostatics.  The next section describes the
traditional, charge-sign symmetric continuum model, our modifications
to incorporate asymmetric response, and a boundary-integral approach
for solving the NLBC model. Section~\ref{sec:results} describes a
MATLAB implementation and presents computational results for
single-atom ions as well as large spheres approximating full-sized
proteins.  The paper concludes in Section~\ref{sec:discussion} with a
summary and discussion.

\psection{THEORY}\label{sec:theory}

\psubsection{Symmetric Continuum Model and Boundary-Integral Method}

We consider a protein in an infinite dilute electrolyte solution.  In
the exterior region, which we label $I$, the potential obeys the
linearized Poisson--Boltzmann equation $\nabla^2 \varphi = \kappa^2
\varphi$ where $\kappa$ represents the inverse Debye screening
length~\cite{Sharp90}, and the dielectric constant is labeled
$\epsilon_I$ (often taken to be 80, approximately that of bulk water).
A thin shell of ion-free solvent separates the electrolyte solution
from the protein.  This region $II$, labeled the Stern or
ion-exclusion layer, is a few Angstroms in width and simplistically
models the finite size of the ions dissolved in the electrolyte.  Here
the potential obeys the Laplace equation and the dielectric constant
is $\epsilon_{II}$ (usually $\epsilon_{II} = \epsilon_I$).  The
protein interior, labeled $III$, is a low-dielectric medium
($\epsilon_{III}$ is usually between 2 and 8) containing $N$ discrete
point charges, and the potential satisfies a Poisson equation
$\nabla^2 \varphi(r) = -\sum_{i=1}^{N} q_i \delta(r-r_i)$ where $q_i$
and $r_i$ specify the $i$th charge.  The boundary $a$ separates the
protein region $III$ and the Stern layer $II$, and the boundary $b$
separates the Stern layer from the electrolyte solvent $I$.  The potential
is assumed to decay to zero suitably fast as $r\rightarrow \infty$, and the
boundary conditions are
\begin{align}
  \varphi_{III}(\mathbf{r}_a) & = \varphi_{II}(\mathbf{r}_a)\\
\epsilon_{III}\frac{\partial \varphi_{III}(\mathbf{r}_a)}{\partial n} & = \epsilon_{II}\frac{\partial \varphi_{II}(\mathbf{r}_a)}{\partial n} \label{eq:smbc-2}\\
  \varphi_{II}(\mathbf{r}_b) & = \varphi_{I}(\mathbf{r}_b)\\
\epsilon_{II}\frac{\partial \varphi_{II}(\mathbf{r}_b)}{\partial n} & = \epsilon_{I}\frac{\partial \varphi_{I}(\mathbf{r}_b)}{\partial n}.\label{eq:smbc-4}
  \end{align}
The flux conditions Eqs.~\ref{eq:smbc-2} and~\ref{eq:smbc-4} are what
we call the standard Maxwell boundary conditions (SMBC).  By applying
Green's theorem in these three regions and taking suitable limits as
the field point approaches each region's bounding surface or surfaces,
we obtain the coupled BIE system~\cite{Yoon90,Altman09,Cooper13}:
\begin{align}
\begin{sbmatrix}{cccc}
  (\frac{1}{2}I + K_{III,aa}) & -G_{III,aa} & 0 & 0\\
  (\frac{1}{2}I - K_{II,aa}) & +G_{II,aa}\frac{\epsilon_{III}}{\epsilon_{II}} & +K_{II,ab} & -G_{II,ab}\\
  -K_{II,ba} & +G_{II,ba}\frac{\epsilon_{III}}{\epsilon_{II}} & (\frac{1}{2}I+K_{II,bb}) & -G_{II,bb}\\
  0 & 0 & (\frac{1}{2}I-K_{I,bb}) &+G_{I,bb}\frac{\epsilon_{II}}{\epsilon_{I}}
  \end{sbmatrix}
\begin{sbmatrix}{c}
  \varphi_{III}(\mathbf{r}_a) \\
  \frac{\partial \varphi_{III}}{\partial n}(\mathbf{r}_a)\\
  \varphi_{II}(\mathbf{r}_b) \\
  \frac{\partial \varphi_{II}}{\partial n}(\mathbf{r}_b)
  \end{sbmatrix}
=
\begin{sbmatrix}{c}
\sum q_i G_{III} \\ 0 \\ 0 \\ 0
  \end{sbmatrix},
  \end{align}
where $G_{X,ij}$ and $K_{X,ij}$ represent the single- and double-layer
boundary-integral operators associated with the Green's function of
region $X$ that map from a distribution on boundary $j$ to the
potential on boundary $i$.

\psubsection{Renormalized Nonlinear Boundary Condition Model}

Our original work on the NLBC model employed only a single interface,
the protein solvent--solute boundary $a$.  For consistency, the
regions it separates are still labeled $III$ (solute) and $II$
(solvent), and instead of Eq.~\ref{eq:smbc-2} we have the nonlinear
flux boundary condition
\begin{align}
  f(E_n(\mathbf{r}_a^{-})) \frac{\partial \varphi_{III}(\mathbf{r}_a)}{\partial n} & = \left(1 + f(E_n(\mathbf{r}_a^-))\right) \frac{\partial \varphi_{II}(\mathbf{r}_a)}{\partial n},\label{eq:nlbc-2}
\end{align}
where the field-dependent nonlinear function $f$ is
\begin{align}
f(E_n) = \frac{\epsilon_{III}}{\epsilon_{II}-\epsilon_{III}} - \alpha \tanh(\beta E_n - \gamma) - \alpha \tanh(-\gamma).\label{eq:nlbc-f}
\end{align}
The first term of Eq.~\ref{eq:nlbc-f} represents the SMBC, and the
last term ensures that the NLBC model recovers the standard model for
weak electric fields, i.e. as $E_n\rightarrow 0$.  As shown in our
earlier work, the NLBC has only three free parameters~$\alpha$,
$\beta$, and $\gamma$, which can be parameterized robustly against
atomistic calculations.  Numerical simulations using the NLBC model
showed excellent agreement with atomistic
results~\cite{Bardhan14_asym}.  However, outside the solute, the
potentials generated by this model fail to satisfy Gauss's law, as can
be seen by considering a Born ion, i.e. a sphere with central charge.
In particular, solutions satisfying SMBC automatically satisfy Gauss's
law, whereas the NLBC cannot simultaneously satisfy Gauss's law and
provide asymmetric response.  This deficiency of central importance
for problems involving multiple solute molecules, e.g. protein-drug
binding; we propose to solve it by including a compensating charge
distribution a few Angstroms away from the solute--solvent interface,
at the Stern layer.  This compensating charge ensures that Gauss's law
is satisfied outside the second boundary.  The modified boundary
condition at $b$ ensures that the model recovers the expected
macroscopic dielectric behavior far from the protein surface, and is
written
\begin{align}
 \left(-\frac{\sum
    q_i}{\epsilon_I}\right) \frac{\partial
    \varphi_{II}(\mathbf{r}_b)}{\partial n} & = \left(\int_b
  \frac{\partial \varphi_{II}(\mathbf{r}_b)}{\partial n}dA \right)
  \frac{\partial \varphi_{I}(\mathbf{r}_b)}{\partial
    n}.\label{eq:nlbc-4}
  \end{align}
Defining $s_1$ and $s_2$ so that Eq.~\ref{eq:nlbc-4} can be written
$s_1 \frac{\partial \varphi_{II}}{\partial n} = s_2 \frac{\partial
  \varphi_I}{\partial n}$, the system of coupled BIEs is
\begin{align}
\begin{sbmatrix}{cccc}
  (\frac{1}{2}I + K_{III,aa}) & -G_{III,aa} & 0 & 0\\
  (\frac{1}{2}I - K_{II,aa}) & +G_{II,aa}\frac{f}{1+f} & +K_{II,ab} & -G_{II,ab}\\
  -K_{II,ba} & +G_{II,ba}\frac{f}{1+f} & (\frac{1}{2}I+K_{II,bb}) & -G_{II,bb}\\
  0 & 0 & (\frac{1}{2}I-K_{I,bb}) &+G_{I,bb}\frac{s_1}{s_2}
  \end{sbmatrix}
\begin{sbmatrix}{c}
  \varphi_{III}(\mathbf{r}_a) \\
  \frac{\partial \varphi_{III}}{\partial n}(\mathbf{r}_a)\\
  \varphi_{II}(\mathbf{r}_b) \\
  \frac{\partial \varphi_{II}}{\partial n}(\mathbf{r}_b)
  \end{sbmatrix}
=
\begin{sbmatrix}{c}
\sum q_i G_{III} \\ 0 \\ 0 \\ 0
  \end{sbmatrix}.
  \end{align}
Using Green's theorem again to determine the reaction potential
$\varphi^{reac}(q)$ in the protein due to the solvent, the
electrostatic charging free energy is then calculated as $\Delta G =
\frac{1}{2} q^T \varphi^{reac}(q) + \varphi^{static}\sum q_i$, where
we have modeled the static (interface) potential as a constant, see
e.g.~\cite{Ashbaugh00,Cerutti07,Bardhan12_asymmetry} (following our
previous work, we model $\varphi^{static} = 10.7$~kcal/mol/$e$).  In
contrast, the standard Poisson theory gives an energy $\Delta G =
\frac{1}{2} q^T L q$ where $L$ is a symmetric negative definite
operator.  

\psection{COMPUTATIONAL
  RESULTS}\label{sec:results}

\psubsection{Numerical Implementation} 

The full MATLAB code to reproduce the calculations and figures in this
paper are freely available online at
\url{http://www.bitbucket.org/jbardhan/piers15-code}.  Our
boundary-element method from earlier work was extended from using only
point-based discretizations of the relevant boundaries and unknown
surface distributions, to use triangular boundary elements with
piecewise-constant basis functions and centroid collocation.
Following our earlier work, we use Picard iteration to solve the
nonlinear BIE problem.  Calculations on spherical molecules used the
earlier point-based implementation for verification against earlier
results.  Calculations for ellipsoids used triangular meshes derived
from the mesh obtained from MATLAB's \texttt{ellipsoid} command, which
takes as input the ellipsoid semi-axis lengths and a number $n$ of
subdivisions, and returns three $(n+1)$ by $(n+1)$ matrices (for the
$x$, $y$, and $z$ coordinates of the mesh vertices), which define
planar quadrilaterals and triangles by subdividing the ellipsoid in
angular coordinates (lines of latitude and longitude).  By iterating
over the quadrilaterals and subdividing, we obtain a triangular mesh
suitable for our existing MATLAB implementation of the
boundary-element method.


The ellipsoidal shape approximations are calculated using standard
methods~\cite{Taylor83,Sigalov06}: the molecule, e.g. a protein, is
defined as a set of atoms which are defined as spheres at specified
locations and with specified radii, and with each possessing a single
point charge at its center. The $N$ atomic positions are represented
by $\mathbf{r}_i = [x_i; y_i; z_i]$ and the radii are
$[a_1,a_2,....,a_N]$. In order to estimate protein shape as an
ellipsoid, we translate the molecule so its center of mass is at the
origin, where the ``mass'' of the molecule is modeled as proportional
to the sum of the atomic volumes, i.e. $M=\sum_{i=1}^{N}(a_i^3)$.  The
center of mass is defined as $\mathbf{r}_c =
(M^{-1})\sum_{i=1}^{N}(a_i^3)(\mathbf{r}_i)$, and so we translate the
atoms according to $\mathbf{r}_i'=\mathbf{r}_i - \mathbf{r}_c$.
Dropping the prime and referring only to the translated coordinates,
the components of the molecule's inertia tensor $I$ are calculated as
\begin{align}
I_{11}&=\sum_{i=1}(m_i(y_i^2+z_i^2+\frac{2}{5}a_i^2))\\
I_{22}&=\sum_{i=1}(m_i(x_i^2+z_i^2+\frac{2}{5}a_i^2))\\
I_{33}&=\sum_{i=1}(m_i(x_i^2+y_i^2+\frac{2}{5}a_i^2))\\
I_{12}&=\sum_{i=1}(m_ix_iy_i)\\
I_{13}&=\sum_{i=1}(m_ix_iz_i)\\
I_{23}&=\sum_{i=1}(m_iy_iz_i),
  \end{align}
and by symmetry $I_{21}=I_{12}$, $I_{31}=I_{13}$, and $I_{32}
=I_{23}$.  The principal moments of inertia of the molecule are the
eigenvalues of $I$.  Choosing $I_{xx}$, $I_{yy}$, and $I_{zz}$ such
that $I_{xx} \leq I_{yy} \leq I_{zz}$, we find the semi-axes $a$, $b$,
and $c$ of an ellipsoid that has the same weight $M$ and principal
moments of inertia~\cite{Taylor83,Sigalov06}.  This leads to
$a=\sqrt{\frac{5}{2M}(-I_{xx} + I_{yy} + I_{zz})}$,
$b=\sqrt{\frac{5}{2M}(I_{xx} - I_{yy} + I_{zz})}$, and
$c=\sqrt{\frac{5}{2M}(I_{xx} + I_{yy} - I_{zz})}$.  This ellipsoid is
the simplest anisotropic approximation to the shape of the
bio-molecule under consideration.

The results in Figure~\ref{fig:charges-in-spheres}(a) represent
continuum-model calculations of monovalent ions ($+1e$ and $-1e$
charges) as a function of ion radius.  For comparison, reference data
for biologically relevant ions obtained from all-atom molecular
dynamics (MD) calculations~\cite{Bardhan14_asym} are plotted as
symbols.  The renormalized NLBC (thick curves) clearly fit the MD
results better than the original NLBC model (thin curves), even though
no new parameters have been introduced.  The symmetric Poisson model
is substantially incorrect; the most frequent approaches for ions
involve adjusting atomic radii on an atom by atom basis, but this is
not possible for multi-atom solutes~\cite{Corbeil10,Mukhopadhyay14}.
Figure~\ref{fig:charges-in-spheres}(b) plots the deviations between
the new NLBC model and a Poisson model that involves sign-symmetric
dielectric response, but incorporates the static potential
contribution: that is, the results plot the effects of asymmetry in
the dielectric response at the molecule surface.  The sample problems
in this figure are a surface charge in a sphere of varying radius
(1.5~\AA~from the surface), or a buried charge at its center.  As
expected, the buried charge experiences essentially symmetric response
once it is more than a few Angstroms from the interface.  In contrast,
the surface charge experiences a surprisingly large asymmetry that is
essentially constant even as the sphere radius increases.  The
magnitude of this asymmetry is in line with previous MD
calculations~\cite{Bardhan12_asymmetry}, and the persistence of this
large deviation from standard models, even for protein-sized
molecules, suggests that including accurate models of asymmetry should
be a main goal in developing improved continuum theory.
\begin{figure}[ht!]
  \centering \begin{tabular}{cc}
    \resizebox{3.0in}{!}{\includegraphics{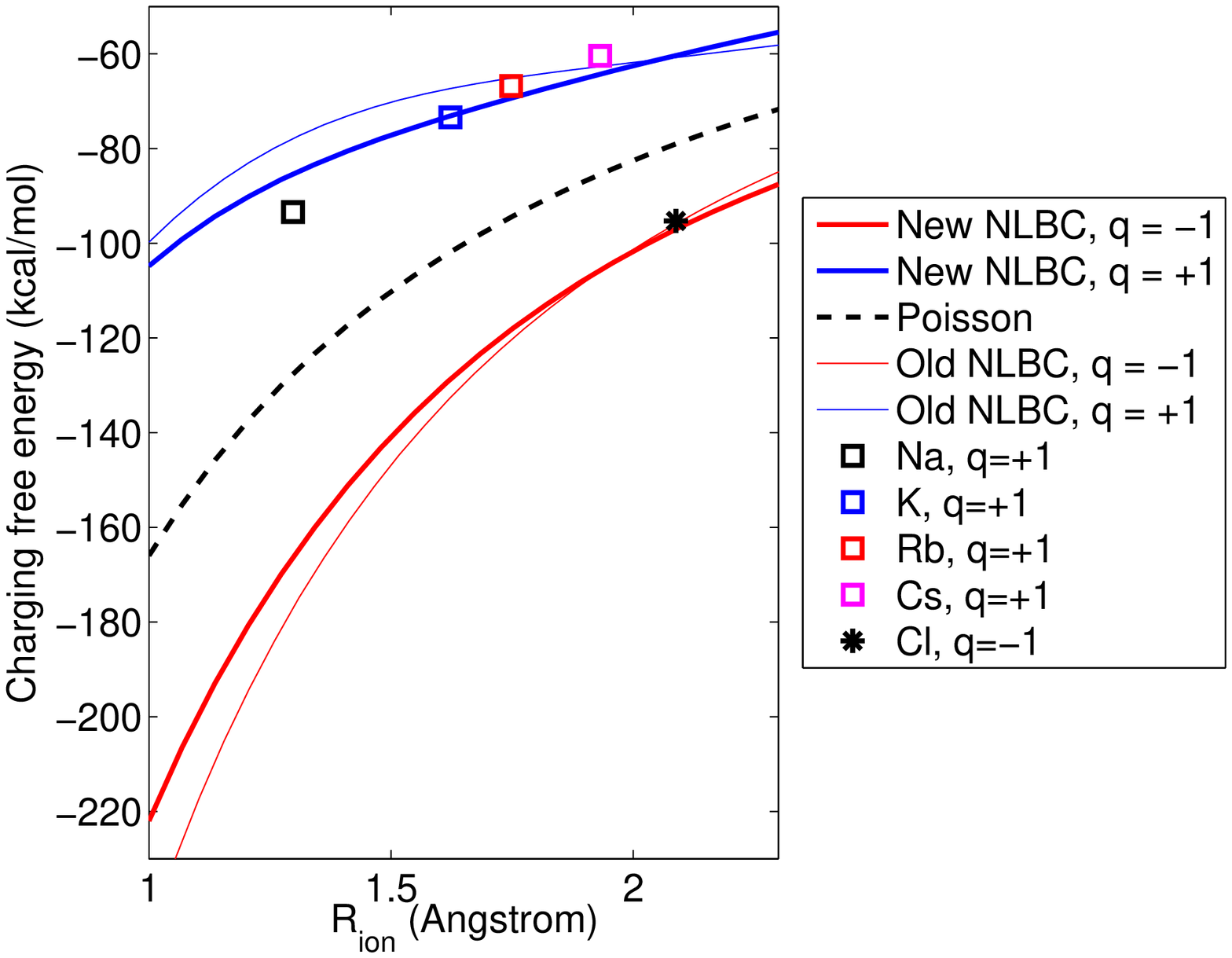}}
    &
    \resizebox{3.0in}{!}{\includegraphics{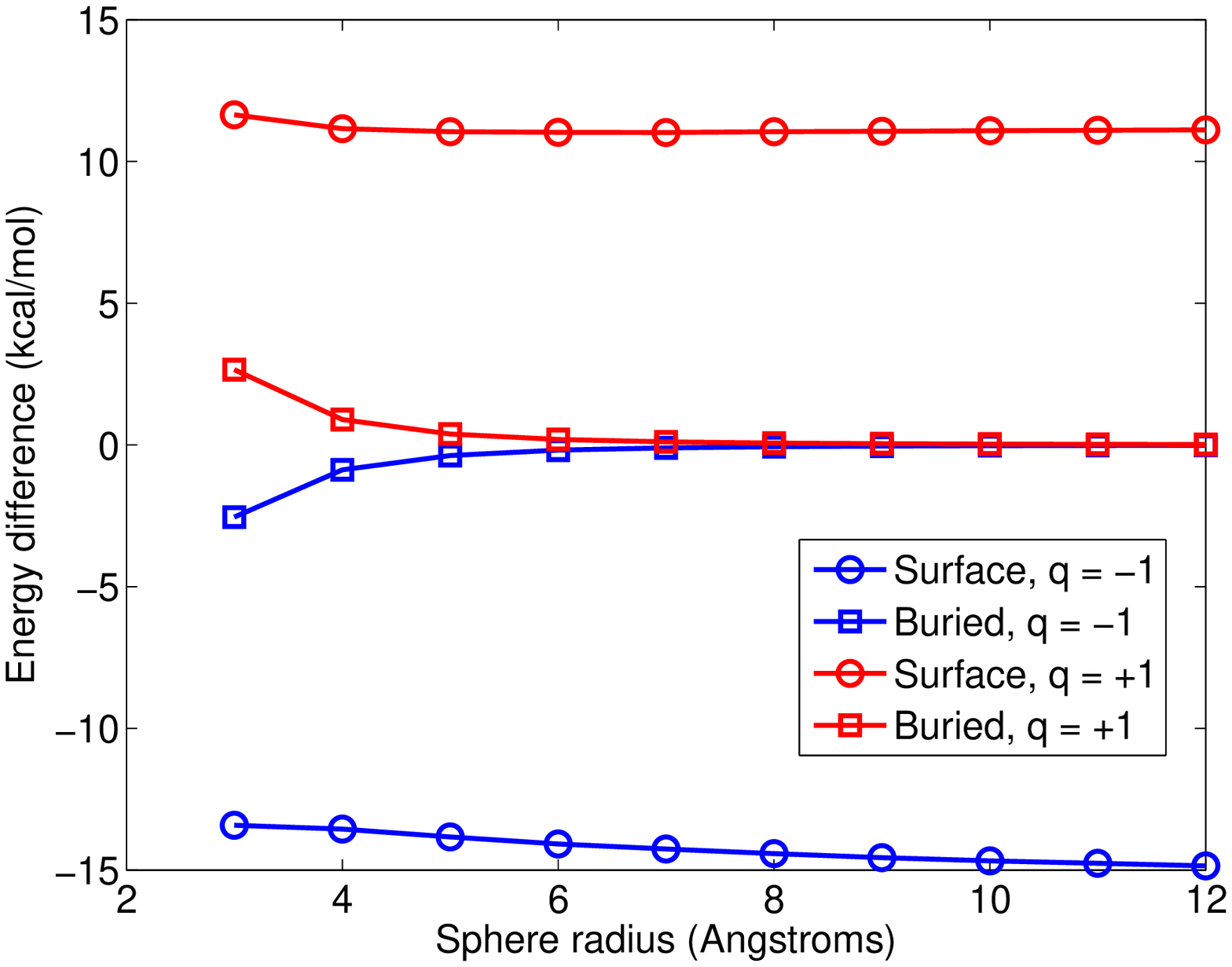}}
    \\
    (a) & (b)
    \end{tabular}
  \caption{Charging free energies for single charges in spheres.  (a)
    Common monovalent ions, calculated using all-atom molecular
    dynamics (symbols)~\cite{Bardhan14_asym}, standard Poisson theory
    (black dashed line), the original NLBC model~\cite{Bardhan14_asym}
    (thin lines), and the new NLBC model presented in this work (thick
    lines).  (b) Charging free energy for a single charge in a sphere,
    as a function of sphere radius, charge value, and charge location.
    Buried charges are situated at the sphere center; surface charges
    are located 1.5~\AA~from the sphere
    surface.}\protect\label{fig:charges-in-spheres} \end{figure}

\psection{DISCUSSION}\label{sec:discussion}

In this work we have presented a continuum-electrostatic model for
molecular solvation that models biologically important hydration
asymmetry--a fundamentally non-continuum phenomenon--using an
effective nonlinear boundary condition (NLBC).  The NLBC replaces the
traditional Maxwell boundary condition that enforces continuity of the
electric flux at the interface, and the present model improves on our
original work~\cite{Bardhan14_asym} by ensuring that Gauss's law holds
outside of the first shell of solvent molecules surrounding molecular
solutes.  The use of an effective boundary condition at the
molecule--solvent interface represents a new frontier in biomolecular
modeling, and was inspired by a long history of approximate boundary
conditions in electromagnetic theory, e.g.~\cite{Senior_Volakis} and
rarefied gases~\cite{Maxwell1878,Smoluchowski1898}.  We hope that the
present work will encourage experts in the electromagnetics community
to contribute their insights and experience to support deeper
understanding of molecular electrostatics, whether in improving
models, analyzing their implications, or solving them numerically.

Our NLBC model is the first asymmetric theory that uses actual Poisson
continuum theory, and in the simple case of a single ion, reproduces
empirical and semi-empirical theories developed over decades of
research into electrostatic
asymmetry~\cite{Latimer39,Corbeil10,Mukhopadhyay14}.  We note that the
present model can treat dilute electrolyte solutions using the
linearized Poisson--Boltzmann equation~\cite{Sharp90,Bardhan12_review}
simply by modeling the Green's function of the outermost region with
the LPBE Green's function $\frac{\exp(-\kappa | r-r'|)}{|r-r'|}$
instead of the $\frac{1}{|r-r'|}$ of the Laplace Green's function.  To
demonstrate the new model's accuracy, we have calculated the
electrostatic charging free energies for single-atom ions and compared
our results to more accurate, and much more computationally demanding,
atomistic molecular dynamics calculations that include thousands of
explicit water molecules.  The results for ions in Figure~1 indicate
that the new model exhibits substantially better accuracy than the
original, and the enforcement of Gauss's law is in fact even more
important when modeling electrolyte solutions using the LPBE (results
not shown).  Calculations of amino-acid charging free energies, and
the differences due to protonation or deprotonation, illustrate that
the magnitude of asymmetric response decays rapidly with an atomic
charge's distance from the solute--solvent
interface~\cite{Bardhan12_asymmetry,Ashbaugh00,Rajamani04}.  Our
numerical calculations employed boundary-integral methods with simple
model geometries, such as spheres for the monatomic ions and
ellipsoids to model the amino acids.  These ellipsoids represent
simple shape approximations~\cite{Taylor83,Sigalov06} and we expect
that they will be useful for fast approximate calculations such as in
implicit-solvent molecular
dynamics~\cite{Bardhan08_BIBEE,Bardhan10,Bardhan11_Knepley,Bardhan12_Knepley_ellipsoidal}.
Calculations for atomistic models of large molecules such as proteins
will require fast, parallel boundary-element method
solvers~\cite{Altman09,Cooper13}, and implementation of such software
represents an area of ongoing work.

\vspace{0.5cm}

\textbf{Mailing address:}\\ Jaydeep P. Bardhan\\ Department of
Mechanical and Industrial Engineering\\ 334 Snell Engineering
Center\\ 360 Huntington Avenue\\ Boston, MA 02115\\ \\ \textbf{Phone:}
617-373-7260\\ \textbf{Fax:} 617-373-2921\\ \\ \\ \textbf{Session:}
Advanced Mathematical and Computational Methods in Electromagnetic
Theory and Their Applications\\ \textbf{Session organizers:} Georgi
Nikolev Georgiev and Mariana Nikolova Georgieva-Grosse\\ \textbf{Oral
  presentation preferred}

\ack
MGK was partially supported by the U.S. Department of Energy,
Office of Science, Advanced Scientific Computing Research, under
Contract DE-AC02-06CH11357, and also NSF Grant OCI-1147680. JPB has
been supported in part by the National Institute of General Medical
Sciences (NIGMS) of the National Institutes of Health (NIH) under
award number R21GM102642.

\bibliographystyle{unsrt}
\bibliography{implicit-review}

\begin{thebibliography}{10}

\bibitem{Sharp90}
K.~A. Sharp and B.~Honig.
\newblock Electrostatic interactions in macromolecules: Theory and
  applications.
\newblock {\em Annu. Rev. Biophys. Bio.}, 19:301--332, 1990.

\bibitem{Bardhan12_review}
J.~P. Bardhan.
\newblock Biomolecular electrostatics---{I} want your solvation (model).
\newblock {\em Computational Science and Discovery}, 5:013001, 2012.

\bibitem{Beglov96}
D.~Beglov and B.~Roux.
\newblock Solvation of complex molecules in a polar liquid: an integral
  equation theory.
\newblock {\em Journal of Chemical Physics}, 104(21):8678--8689, 1996.

\bibitem{Hildebrandt04}
A.~Hildebrandt, R.~Blossey, S.~Rjasanow, O.~Kohlbacher, and H.-P. Lenhof.
\newblock Novel formulation of nonlocal electrostatics.
\newblock {\em Phys. Rev. Lett.}, 93:108104, 2004.

\bibitem{Kirkwood34}
J.~G. Kirkwood.
\newblock Theory of solutions of molecules containing widely separated charges
  with special application to zwitterions.
\newblock {\em J. Chem. Phys.}, 2:351, 1934.

\bibitem{Phillips05}
J.~C. Phillips, R.~Braun, W.~Wang, J.~Gumbart, E.~Tajkhorshid, E.~Villa,
  C.~Chipot, R.~D. Skeel, L.~Kale, and K.~Schulten.
\newblock Scalable molecular dynamics with {NAMD}.
\newblock {\em J. Comput. Chem.}, 26:1781--1802, 2005.

\bibitem{Roux99}
B.~Roux and T.~Simonson.
\newblock Implicit solvent models.
\newblock {\em Biophys. Chem.}, 78:1--20, 1999.

\bibitem{Ashbaugh00}
H.~S. Ashbaugh.
\newblock Convergence of molecular and macroscopic continuum descriptions of
  ion hydration.
\newblock {\em The Journal of Physical Chemistry B}, 104(31):7235--7238, 2000.

\bibitem{Rajamani04}
S.~Rajamani, T.~Ghosh, and S.~Garde.
\newblock Size dependent ion hydration, its asymmetry, and convergence to
  macroscopic behavior.
\newblock {\em J. Chem. Phys.}, 120:4457, 2004.

\bibitem{Mobley08}
D.~L. Mobley, K.~A. Dill, and J.~D. Chodera.
\newblock Treating entropy and conformational changes in implicit solvent
  simulations of small molecules.
\newblock {\em J. Phys. Chem. B}, 112:938--946, 2008.

\bibitem{Bardhan12_asymmetry}
J.~P. Bardhan, P.~Jungwirth, and L.~Makowski.
\newblock Affine-response model of molecular solvation of ions: Accurate
  predictions of asymmetric charging free energies.
\newblock {\em J. Chem. Phys.}, 137:124101, 2012.

\bibitem{Bardhan11_pka}
J.~P. Bardhan.
\newblock Nonlocal continuum electrostatic theory predicts surprisingly small
  energetic penalties for charge burial in proteins.
\newblock {\em J. Chem. Phys.}, 135:104113, 2011.

\bibitem{Bardhan13_nonlocal_review}
J.~P. Bardhan.
\newblock Gradient models in molecular biophysics: progress, challenges,
  opportunities.
\newblock {\em Journal of Mechanical Behavior of Materials}, 22:169--184, 2013.

\bibitem{Bardhan14_asym}
J.~P. Bardhan and M.~G. Knepley.
\newblock Modeling charge-sign asymmetric solvation free energies with
  nonlinear boundary conditions.
\newblock {\em J. Chem. Phys.}, 141:131103, 2014.

\bibitem{Dogonadze74}
R.~R. Dogonadze and A.~A. Kornyshev.
\newblock Polar solvent structure in the theory of ionic solvation.
\newblock {\em J. Chem. Soc. Faraday Trans. 2}, 70:1121--1132, 1974.

\bibitem{Fedorov07}
M.~V. Fedorov and A.~A. Kornyshev.
\newblock Unravelling the solvent response to neutral and charged solutes.
\newblock {\em Molecular Physics}, 105:1--16, 2007.

\bibitem{Latimer39}
W.~M. Latimer, K.~S. Pitzer, and C.~M. Slansky.
\newblock The free energy of hydration of gaseous ions, and the absolute
  potential of the normal calomel electrode.
\newblock {\em J. Chem. Phys.}, 7:108--112, 1939.

\bibitem{Grossfield05}
A.~Grossfield.
\newblock Dependence of ion hydration on the sign of the ion's charge.
\newblock {\em J. Chem. Phys.}, 122:024506, 2005.

\bibitem{Corbeil10}
C.~R. Corbeil, T.~Sulea, and E.~O. Purisima.
\newblock Rapid prediction of solvation free energy. 2. {The} first-shell
  hydration {(FiSH)} continuum model.
\newblock {\em J. Chem. Theory Comput.}, 6:1622--1637, 2010.

\bibitem{Mukhopadhyay14}
A.~Mukhopadhyay, B.~H. Aguilar, I.~S. Tolokh, and A.~V. Onufriev.
\newblock Introducing charge hydration asymmetry into the {Generalized Born}
  model.
\newblock {\em J. Chem. Theory Comput.}, 10:1788--1794, 2014.

\bibitem{Cerutti07}
D.~S. Cerutti, N.~A. Baker, and J.~A. {McCammon}.
\newblock Solvent reaction field potential inside an uncharged globular
  protein: a bridge between implicit and explicit solvent models?
\newblock {\em J. Chem. Phys.}, 127:155101, 2007.

\bibitem{Kathmann11}
S.~M. Kathmann, I-F.~W. Kuo, C.~J. Mundy, and G.~K. Schenter.
\newblock Understanding the surface potential of water.
\newblock {\em J. Phys. Chem. B}, 115:4369--4377, 2011.

\bibitem{Purisima09}
E.~O. Purisima and T.~Sulea.
\newblock Restoring charge asymmetry in continuum electrostatic calculations of
  hydration free energies.
\newblock {\em J. Phys. Chem. B}, 113:8206--8209, 2009.

\bibitem{Altman09}
M.~D. Altman, J.~P. Bardhan, J.~K. White, and B.~Tidor.
\newblock Accurate solution of multi-region continuum electrostatic problems
  using the linearized {Poisson}--{Boltzmann} equation and curved boundary
  elements.
\newblock {\em J. Comput. Chem.}, 30:132--153, 2009.

\bibitem{Bardhan10}
J.~P. Bardhan.
\newblock Rapid bounds on electrostatic energies using diagonal approximations
  of boundary-integral equations.
\newblock {\em Progress in Electromagnetics Research Symposium (PIERS)}, 2010.

\bibitem{PIERS15_repository}
J.~P. Bardhan, D.~Tejani, N.~Wieckowski, A.~Ramaswamy, and M.~G. Knepley.
\newblock Public git repository containing all source code and data to
  reproduce the figures in this paper.
\newblock https://bitbucket.org/jbardhan/piers15-code.

\bibitem{Yoon90}
B.~J. Yoon and A.~M. Lenhoff.
\newblock A boundary element method for molecular electrostatics with
  electrolyte effects.
\newblock {\em J. Comput. Chem.}, 11(9):1080--1086, 1990.

\bibitem{Cooper13}
C.~D. Cooper, J.~P. Bardhan, and L.~A. Barba.
\newblock A biomolecular electrostatics solver using {Python}, {GPUs} and
  boundary elements that can handle solvent-filled cavities and {Stern} layers.
\newblock {\em Comput. Phys. Commun.}, 185:720--729, 2013.

\bibitem{Taylor83}
W.~R. Taylor, J.~M. Thornton, and W.~G. Turnell.
\newblock An ellipsoidal approximation of protein shape.
\newblock {\em J. Mol. Graph.}, 1:30--38, 1983.

\bibitem{Sigalov06}
G.~Sigalov, A.~Fenley, and A.~Onufriev.
\newblock Analytical electrostatics for biomolecules: {Beyond} the generalized
  {Born} approximation.
\newblock {\em J. Chem. Phys.}, 124(124902), 2006.

\bibitem{Senior_Volakis}
T.~B.~A. Senior and J.~L. Volakis.
\newblock {\em Approximate boundary conditions in electromagnetics}.
\newblock IEEE, London, 1995.

\bibitem{Maxwell1878}
James~Clerk Maxwell.
\newblock On stresses in rarefied gases arising from inequalities of
  temperature.
\newblock {\em Proceedings of the Royal Society of London}, 27:304--308, 1878.

\bibitem{Smoluchowski1898}
Marian von Smolan~Smoluchowski.
\newblock {\"Uber} w\"armeleitung in verd\"unnten gasen.
\newblock {\em Annalen der Physik}, 300(1):101--130, 1898.

\bibitem{Bardhan08_BIBEE}
J.~P. Bardhan.
\newblock Interpreting the {Coulomb}-field approximation for
  {Generalized}-{Born} electrostatics using boundary-integral equation theory.
\newblock {\em J. Chem. Phys.}, 129(144105), 2008.

\bibitem{Bardhan11_Knepley}
J.~P. Bardhan and M.~G. Knepley.
\newblock Mathematical analysis of the boundary-integral based electrostatics
  estimation approximation for molecular solvation: Exact results for spherical
  inclusions.
\newblock {\em J. Chem. Phys.}, 135:124107, 2011.

\bibitem{Bardhan12_Knepley_ellipsoidal}
J.~P. Bardhan and M.~G. Knepley.
\newblock Computational science and re-discovery: open-source implementation of
  ellipsoidal harmonics for problems in potential theory.
\newblock {\em Computational Science and Discovery}, 5:014006, 2012.

\end{thebibliography}

\end{paper}

\end{document}